# New theoretical results on the Universe Structure.

F.A. Hovsepian (Moscow)

## 1. Introduction

This paper discusses an ordinary homogeneous differential equation of the second order with constant real-valued coefficients. It is one of the three equations describing the motion of a celestial body of natural origin in the Euclidean space. The y(t) solution of this equation has to satisfy additional conditions which are very significant in the author's opinion and which, so far, have escaped attention of the researchers.

*The first of the conditions* stems from the analytical research results of numerous scientists and many years of astronomic observations confirming the stability of the celestial body motion. Since stable motion in nature is rotational, normally consideration is given to a bounded second-order equation with the coefficient of the first derivative y(t) equal to zero. This paper discusses only an asymptotically stable solution of the y(t) equation which is needed from the theoretical point of view as well, since in what follows the solution is Fourier's expansion.

*The second condition* is based on the fact that nature does not know of such notions as left or right, up or down. Consequently, y(t) for $t \geq 0$ is reflected with respect to $t \leq 0$, i.e. an even function is obtained $v(\tau) \to 0$ for $\tau \to \pm \infty$, and the continuous differentiability of function $v(\tau)$ for $\tau = 0$ is maintained via two starting conditions that the researcher has at his disposal. Consequently, function $v(\tau)$ is a solution of two second order equations which differ one from the other by the first derivative sign.

It is well known that Fourier's transform of $v(\tau)$ is

$$V(\omega) = \frac{1}{2\pi} \int_{-\infty}^{\infty} v(\tau)\exp(-i\omega\tau)d\tau.$$

In the case considered $V(\omega)$ has several very important properties. First, $V(\omega)$ due to the even $v(\tau)$ is a real valued function. Second, function $V(\omega) = V(-\omega) \geq 0$ for all $\omega$, i.e. this function is the probability density of a random value. Hence, the expansion of the function

$$v(\tau) = 2\int_{0}^{\infty} V(\omega)\cos\omega\tau \, d\omega$$

has an important physical meaning which can be ascribed to it based on the known Khinchine's theorem [1]: $v(\tau)$ – is an autocorrelation function of a continuous stationary random process $\xi(t)$. Consequently, the argument $\tau$ of this function is a time interval between sections of this process at arbitrary times $t_1$ and $t_2$:

$$\tau = t_1 - t_2.$$

Note that in a conventional solution y(t), t = 0 is a point, however, in the solution v($\tau$) for $\tau$ = 0 we have current time t, which lasts from minus to plus infinity. This is one of the two significant results of this paper and the proof is given in theorem 1.

Theorem 2 has an auxiliary role and is needed for the future research. It proves that the function v($\tau$) can be twice differentiated under the integral with respect to $\tau$ as a parameter. Substituting v($\tau$) in the second order equation for $\tau \geq 0$ and having differentiated twice this function under the integral we obtain an identity with several remarkable properties.

*The first property* of the identity implies that a change in the scale of the integration variable, i.e. a substitution

$$\omega = \frac{\omega_1}{\alpha} \ (\alpha > 0),$$

results in a change of the scale for $\tau$ so that the product stays unchanged:

$$\omega\tau = \frac{\omega_1}{\alpha}\alpha\tau_1 .$$

This is proved in theorem 3.

*The second property* of this identity consists in the fact that a change in the sign of the frequency $\omega$ changes the sign of $\tau$. Consequently, in

$$\frac{dv(\tau)}{d\tau} = \int_0^\infty \omega V(\omega)\sin \omega\tau d\omega$$

when expanding limits of integration with respect to $\omega$ from $-\infty$ to $\infty$ the function sin$\omega\tau$ «behaves» as an even function, and this makes the integrand odd. This property can be utilized after obtaining through the use of Dirac's $\delta$–functions of the following solution

$$v_m(\tau) = \frac{1}{2} \int_{-\infty}^\infty [\delta_1(\Omega - \omega) + \delta_2(\Omega + \omega)] \ V(\Omega)\cos\Omega\tau d\Omega$$

of the equation in question. For any pair of frequencies $\pm\omega$ we have

$$\frac{dv_m(\tau)}{d\tau} \equiv 0,$$

which is proved in Theorem 4. This condition «by-passes» the first derivative coefficient in the second order equation and we obtain an equation which looks like the one describing oscillations. From the interval ($-\infty, \infty$) through the use of $v_m(\tau)$ oscillations of all frequencies except for are $\omega$ = 0 are identified and, as a result, we have the following situation. On the one hand, the function $v_m(\tau)$ is oscillation cos$\Omega\tau$, in which possible values of the random value $\Omega$ are discrete and equal to $\pm\omega$, i.e. an oscillating body behaves as a particle. On the other hand, however, the random value $\Omega$ can be treated as continuous since all its possible values are frequencies from the probability density spectrum V(ω) (ω ≠ 0), i.e. an oscillating body behaves as a wave.

This corresponds to the corpuscular-wave dualism for radiation introduced in physics by Louis de Broglie. Based on this the present author identifies the oscillations as white light. The result presents dual interest:

    1. It has not been known in physics the nature of light generation. We have an answer to this question now.

    2. The absolute time t has been assumed to be an argument in harmonic oscillations which represent light. The present result refutes this assumption. An argument here is the time interval $\tau$ between sections of a stationary random process $\xi(t)$, which changes radically the idea of the Universe composition and the processes taking place in it. Light is the correlation function of the process $\xi(t)$, on the one hand, and manifestation of Euclidean and stationary Universe, on the other hand.

If the Universe were stationary, it could not have had processes like spreading out of galaxies, observed by astronomers. In paragraph 3, among other things, this issue is also discussed using the example of an equation with a parameter which can assume various values, thus offering a possibility to compare various differential equations.

The author's results of [2, 3] are heavily used in the article.

## 2. Theorems

*Theorem* 1. Let y(t) be a partial solution of an ordinary homogeneous asymptotically stable differential equation with constant real-valued coefficients of the second order with initial conditions

$$y(t)\big|_{t=0} = 1 \text{ and } \frac{dy(t)}{dt}\bigg|_{t=0} = 0.$$

Then the function $v(\tau)$, which consists of y(t) for t ≥ 0 and y(t) evenly extended to t < 0 is a correlation function of a continuous stationary process.

*Proof.* The solution y(t) of the second order differential equation is determined by pair of roots $\rho_1$ and $\rho_2$ considered below.

    1. Let the solution y(t) be determined by a pair of complex-conjugated numbers $\rho_{1,2} = -\chi \pm i\lambda$, i.e.

$$y_1(t) = (A\cos\lambda t + B\sin\lambda t)\exp(-\chi t),$$

which, taking into account the initial conditions, can be rewritten as

$$y_1(t) = (\cos\lambda t + \frac{\chi}{\lambda}\sin\lambda t)\exp(-\chi t).$$

Consider $y_1(t)$ for $t \geq 0$ and substitute $t$ for $-s$, rewriting it as follows

$$y_1^*(s) = y_1(-s).$$

Obtain an even function

$$v_1(t) = y_1^*(t) \text{ for } t \leq 0, \quad v_1(t) = y_1(t) \text{ for } t \geq 0.$$

The Fourier transform $V_1(i\omega)$ of the function $v_1(t)$ is equal to

$$V_1(i\omega) = \frac{1}{2\pi} \int_{-\infty}^{\infty} v_1(t)\exp(-i\omega t)dt = \frac{1}{2\pi}[\int_{-\infty}^{0} y_1^*(t) \exp(-i\omega t)dt + \int_{0}^{\infty} y_1(t) \exp(-i\omega t)dt] =$$

$$= \frac{1}{2\pi}[\int_{0}^{\infty} y_1^*(-s)\exp(i\omega s)ds + \int_{0}^{\infty} y_1(t)\exp(-i\omega t)dt].$$

Since we agreed that $y_1^*(-s) = y_1(s)$, then the integrals in $V_1(i\omega)$ differ only in the sign $\omega$, hence, their sum is a real-valued function which permits rewriting the Fouruir transform of the function $v_1(t)$ as

$$V_1(\omega) = \frac{1}{2\pi} \int_{-\infty}^{\infty} v_1(t)\exp(-i\omega t)dt = \frac{1}{\pi} \int_{0}^{\infty} v_1(t)\cos\omega t\, dt$$

For calculation of $V_1(\omega)$ we use integrals (A.4) and (A.5) from Annex:

$$V_1(\omega) = \frac{1}{\pi} \int_{0}^{\infty} [(\cos\lambda t + \frac{\chi}{\lambda}\sin\lambda t)\exp(-\chi t)]\cos\omega t\, dt =$$

$$= \frac{1}{\pi} \{ \frac{\chi(\chi^2 + \lambda^2 + \omega^2)}{[\chi^2 + (\lambda+\omega)^2][\chi^2 + (\lambda-\omega)^2]} + \frac{\chi}{\lambda} \frac{\lambda(\chi^2 + \lambda^2) - \lambda\omega^2}{[\chi^2 + (\lambda+\omega)^2][\chi^2 + (\lambda-\omega)^2]} \} =$$

$$= \frac{2\chi(\chi^2 + \lambda^2)}{\pi[\chi^2 + (\lambda+\omega)^2][\chi^2 + (\lambda-\omega)^2]} \geq 0$$

for all $\omega \geq 0$.

2. Let the solution $y(t)$ be determined by two different real-valued numbers $\rho_1 = -\chi_1$ and $\rho_2 = -\chi_2$, т.е.

$$y_2(t) = A\exp(-\chi_1 t) + B\exp(-\chi_2 t),$$

which taking into account the initial conditions we write as

$$y_2(t) = \frac{\chi_2}{(\chi_2 - \chi_1)}[\exp(-\chi_1 t) - \frac{\chi_1}{\chi_2}\exp(-\chi_2 t)].$$

In this case too we can obtain an even function $v_2(t)$ as

$$v_2(t) = y_2(-t) = y_2^*(t) \text{ for } t \leq 0, \quad v_2(t) = y_2(t) \text{ for } t \geq 0$$

with bilateral Fourier transform

$$V_2(i\omega) = \frac{1}{2\pi} \int_{-\infty}^{\infty} v_2(t)\exp(-i\omega t)dt = \frac{1}{\pi} \int_{0}^{\infty} v_2(t)\cos\omega t\, dt =$$

$$= \frac{1}{\pi} \int_{0}^{\infty} \frac{\chi_2}{(\chi_2 - \chi_1)}[\exp(-\chi_1 t) - \frac{\chi_1}{\chi_2}\exp(-\chi_2 t)]\cos\omega t\, dt.$$

Using the integral value (A.1) from Annex, we can write

$$V_2(\omega) = \frac{1}{\pi} \frac{\chi_2}{(\chi_2 - \chi_1)} \{\frac{\chi_1}{\chi_1^2 + \omega^2} - \frac{\chi_1}{\chi_2} \frac{\chi_2}{\chi_2^2 + \omega^2}\} = \frac{\chi_2}{\pi(\chi_2 - \chi_1)} \frac{\chi_1(\chi_2^2 + \omega^2) - \chi_1(\chi_1^2 + \omega^2)}{\pi(\chi_1^2 + \omega^2)(\chi_2^2 + \omega^2)} =$$

$$= \frac{\chi_1 \chi_2 (\chi_1 + \chi_2)}{\pi(\chi_1^2 + \omega^2)(\chi_2^2 + \omega^2)} \geq 0$$

for all $\omega \geq 0$.

3. Let the solution y(t) be determined by one real-valued number of double multiplicity $\rho_1 = \rho_2 = -\chi$, т.е.

$$y_3(t) = A\exp(-\chi t) + Bt\exp(-\chi t),$$

which taking into account initial conditions we can write as

$$y_3(t) = \exp(-\chi t) + \chi t\exp(-\chi t).$$

In this case too one can obtain an even function $v_3(t)$ following the above rule. The bilateral Fourier transform of function $v_3(t)$ is equal

$$V_3(i\omega) = \frac{1}{2\pi} \int_{-\infty}^{\infty} v_3(t)\exp(-i\omega t)dt = \frac{1}{\pi} \int_0^{\infty} v_3(t)\cos\omega t\, dt =$$

$$= \frac{1}{\pi} \int_0^{\infty} [\exp(-\chi t) + \chi t\exp(-\chi t)]\cos\omega t\, dt.$$

Using integrals (A.1) and (A.7) from Annex, we write

$$V_3(\omega) = \frac{1}{\pi} \{\frac{\chi}{\chi^2 + \omega^2} + \frac{\chi(\chi^2 - \omega^2)}{(\chi^2 + \omega^2)^2}\} = \frac{\chi(\chi^2 + \omega^2) + \chi(\chi^2 - \omega^2)}{\pi(\chi^2 + \omega^2)^2} = \frac{2\chi^3}{\pi(\chi^2 + \omega^2)^2} \geq 0$$

for all $\omega \geq 0$.

The Fourier transform $V_i(\omega)$ of the function $v_i(\tau)$ (i = 1, 2, 3) for all root types in a second order equation is a non-negative function, consequently, $V_i(\omega)$ is probability density of a random value $\Omega_i$. Since $v_i(\tau)$ is a continuous function and

$$\int_{-\infty}^{\infty} |v_i(\tau)|\, d\tau < \infty,$$

then $v_i(\tau)$, as is known, can be presented in the form of the Fourier integral

$$v_i(\tau) = \int_{-\infty}^{\infty} \exp(i\omega\tau)V_i(\omega)d\omega = 2\int_0^{\infty} V_i(\omega)\cos\omega\tau\, d\omega.$$

According to Khinchin's theorem [1] the function $v_i(\tau)$ for all i = 1, 2, 3 is an autocorrelation function of a continuous stationary process.

The Theorem 1 is proved.

*Theorem* 2. Let there be a function

$$v_i(\tau) = 2\int_0^{\infty} V_i(\omega)\cos\omega\tau\, d\omega, \quad (i = 1, 2, 3) \quad (-\infty < \tau < \infty) \qquad (2.1)$$

which in accordance with theorem 1 is a solution of the differential equation

$$\frac{d^2v}{d\tau^2} + a\frac{dv}{d\tau} + bv = 0. \quad (\tau \geq 0) \qquad (2.2)$$

Then $v_i(\tau)$:

a) twice differentiated under the sign of integration with respect to $\tau$ as a parameter, i.e.

$$\frac{d^j v_i}{d\tau^j} = 2\frac{d^j}{d\tau^j}\int_0^\infty V_i(\omega)\cos\omega\tau\, d\omega = 2\int_0^\infty V_i(\omega)\frac{\partial^j \cos\omega\tau}{\partial\tau^j}d\omega; \quad (j = 1, 2)$$

b) satisfies the identities

$$-\int_0^\infty \omega^2 V_i(\omega)\cos\omega\tau\, d\omega - a\int_0^\infty \omega V_i(\omega)\sin\omega\tau\, d\omega + b\int_0^\infty V_i(\omega)\cos\omega\tau\, d\omega \equiv 0, \quad (\tau \geq 0) \qquad (2.3)$$

*Proof*. The function

$$v_i(\tau) = 2\int_0^\infty V_i(\omega)\cos\omega\tau\, d\omega \qquad (i = 1, 2, 3).$$

can be differentiated under the sign of integration with respect to $\tau$, if all improper integrals resulting from the differentiation converge uniformly. It is known that the improper integral

$$\int_0^\infty f(x, z)dx$$

converges uniformly in each set S of values z, if for any $z \in S$ and any x from the integration interval the inequality $|f(x, z)| \leq g(x)$, holds where $g(x)$ is a function of comparison, whose integral converges:

$$\int_0^\infty g(x)dx < \infty.$$

According to the above the integral

$$\int_0^\infty \frac{\partial^2}{\partial\tau^2} V_i(\omega)\cos\omega\tau\, d\omega = -\int_0^\infty \omega^2 V_i(\omega)\cos\omega\tau d\omega$$

converges uniformly, since $|\omega^2 V_i(\omega)\cos\omega\tau| \leq \omega^2 V_i(\omega)$, and

$$\int_0^\infty \omega^2 V_i(\omega)d\omega < \infty,$$

which is easily checked by direct integration. It is obvious now that the integral

$$\int_0^\infty \frac{\partial}{\partial\tau} V_i(\omega)\cos\omega\tau\, d\omega = -\int_0^\infty \omega V_i(\omega)\sin\omega\tau\, d\omega,$$

also converges uniformly, since

$$|\omega V_i(\omega)\sin\omega\tau| \leq \omega V_i(\omega) \text{ for } \omega \geq 0.$$

The integral

$$\int_0^\infty \omega V_i(\omega)d\omega < \infty,$$

is easily tested by direct integration, and the convergence of this integral proves statement a) of theorem 2.

The solution (2.1) is substituted in equations (2.2) and, after differentiating under the sign of integration with respect to τ as a parameter, and in line with the above proof, we obtain identities (2.3), which proves statement b) of theorem 2.

Theorem 2 is proved completely.

*Theorem* 3. Let there be an identity (2.3) from theorem 2 and let there be a substitution of the integration variable $\omega = \dfrac{\omega_1}{\alpha}$ ($\alpha > 0$).

Then:

1) the coefficients and argument of equation (2.2) are changed

$$\frac{d^2 v}{d\tau_1^2} + \alpha a \frac{dv}{d\tau_1} + \alpha^2 b v = 0, \quad (\tau_1 \geq 0) \qquad (2.4)$$

$$\tau_1 = \frac{\tau}{\alpha};$$

2) the product of frequency ω and argument τ isn't changed, i.e

$$\omega\tau = \frac{\omega_1}{\alpha}\alpha\tau_1 = \omega_1\tau_1; \qquad (\omega_1 > 0, \tau_1 > 0) \qquad (2.5)$$

3) instead of the identity (2.3) one obtains a new identity

$$-\int_0^\infty \omega^2 \overline{V}_i(\omega)\cos\omega\tau_1\, d\omega - \alpha a \int_0^\infty \omega \overline{V}_i(\omega)\sin\omega\tau_1\, d\omega + \alpha^2 b \int_0^\infty \overline{V}_i(\omega)\cos\omega\tau_1\, d\omega \equiv 0, \quad (\tau_1 \geq 0) \quad (2.6)$$

in which

$$\overline{V}_i(\omega) = \frac{1}{\pi} \int_0^\infty \overline{v}_i(\tau_1)\cos\omega\tau_1\, d\tau_1, \qquad \overline{v}_i(\tau_1) = v_i(\tau).$$

*Proof.* Consider an identity (2.3) from theorem 2 when i = 1, i.e.

$$-\int_0^\infty \omega^2 V_1(\omega)\cos\omega\tau\, d\omega - a\int_0^\infty \omega V_1(\omega)\sin\omega\tau\, d\omega + b\int_0^\infty V_1(\omega)\cos\omega\tau\, d\omega \equiv 0, \qquad (2.7)$$

where

$$V_1(\omega) = \frac{1}{\pi}\int_0^\infty v_1(\tau)\cos\omega\tau\, d\tau, \qquad v_1(\tau) = (\cos\lambda\tau + \frac{\chi}{\lambda}\sin\lambda\tau)\exp(-\chi\tau). \quad (\tau \geq 0)$$

Substitute $\omega = \dfrac{\omega_1}{\alpha}$ ($\alpha > 0$) in the function

$$V_1(\omega) = \frac{1}{\pi} \frac{2\chi(\chi^2+\lambda^2)}{[\chi^2+(\lambda+\omega)^2][\chi^2+(\lambda-\omega)^2]} = \frac{2\chi(\chi^2+\lambda^2)}{\pi[\chi^2+(\lambda+\frac{\omega_1}{\alpha})^2][\chi^2+(\lambda-\frac{\omega_1}{\alpha})^2]} =$$

$$= \alpha \frac{2\chi_1(\chi_1^2+\lambda_1^2)}{\pi[\chi_1^2+(\lambda_1+\omega_1)^2][\chi_1^2+(\lambda_1-\omega_1)^2]} = V_1^*(\omega_1),$$

where $\chi_1 = \alpha\chi$, $\lambda_1 = \alpha\lambda$.

After substitution in the identity (2.7) we obtain

$$-\int_0^\infty \omega_1^2 V_1^*(\omega_1) \cos\frac{\omega_1}{\alpha}\tau d\omega_1 - \alpha a \int_0^\infty \omega_1 V_1^*(\omega_1) \sin\frac{\omega_1}{\alpha}\tau d\omega_1 + \alpha^2 b \int_0^\infty V_1^*(\omega_1) \cos\frac{\omega_1}{\alpha}\tau \, d\omega_1 \equiv 0. \quad (2.8)$$

The substitution resulted in the following changes:

– in the function $V_1(\omega)$ prior to the substitution we had parameters $\lambda$ and $\chi$, after the substitution we have in the function $V_1^*(\omega_1)$ new parameters

$$\chi_1 = \alpha\chi, \quad \lambda_1 = \alpha\lambda;$$

– a characteristic equation corresponding to the differential equation (2.2) prior to the substitution was

$$\eta^2 + a\eta + b = 0,$$

while, after the substitution it became

$$\eta^2 + \alpha a\eta + \alpha^2 b = 0.$$

Consequently, the variable substitution resulted in changed values of coefficients a and b in the equation (2.2) and they are now equal to $\alpha a$ and $\alpha^2 b$. These changes are not possible since coefficients in a differential equation are specified. Therefore the argument $\tau$ in $v_1(\tau)$ should also change so that in (2.2) the coefficients remain unchanged. Hence, taking into account the notation in $V_1^*(\omega_1)$ we obtain

$$v_1(\tau) = (\cos\lambda\tau + \frac{\chi}{\lambda}\sin\lambda\tau)\exp(-\chi\tau) = (\cos\alpha\lambda\frac{\tau}{\alpha} + \frac{\chi}{\lambda}\sin\alpha\lambda\frac{\tau}{\alpha})\exp(-\alpha\chi\frac{\tau}{\alpha}) =$$

$$= (\cos\alpha\lambda\tau_1 + \frac{\alpha\chi}{\alpha\lambda}\sin\alpha\lambda\tau_1)\exp(-\alpha\chi\tau_1) = \bar{v}_1(\tau_1),$$

where

$$\tau_1 = \frac{\tau}{\alpha},$$

and the equation (2.2) is written as (2.4). At the same time we obtain, that the frequency $\omega$ and argument $\tau$ are linked with a scale-invariant relationship (2.5):

$$\omega\tau = \frac{\omega_1}{\alpha}\alpha\tau_1.$$

The Fourier transform of the function $\bar{v}_1(\tau_1)$, so similar to $v_1(\tau)$, we obtain

$$\bar{V}_1(\omega) = \frac{1}{\pi}\int_0^\infty \bar{v}_1(\tau_1) \cos\omega\tau_1 \, d\tau_1.$$

Substituting $\overline{V}_1(\omega)$ in (2.4) we obtain a new identity (2.6) in which i = 1.

Theorem 3 is proved for i = 1 in the identity (2.3).

The prove of this theorem is similar for i = 2, 3 n the identity (2.3).

The theorem 3 is proved fully.

*Theorem* 4. Let an identity (2.6) from theorem 3 be given and let the integration limits be extended by $\omega$ from $-\infty$ to $\infty$.

Then the equation (2.4) has a new solution

$$v_m(\tau_1) = \frac{1}{2} \int_{-\infty}^{\infty} [\delta_1(\Omega - \omega) + \delta_2(\Omega + \omega)] \overline{V}_i(\Omega) \cos\Omega\tau_1 d\Omega \qquad (2.9)$$

where $\delta_1(\Omega - \omega)$ and $\delta_2(\Omega + \omega)$ are Dirac's functions, if the frequency $\omega$, the arbitrary parameter $\alpha$ and the coefficient b in (2.4) to link by the relationship

$$\frac{\omega}{\alpha} = \pm\sqrt{b} . \qquad (2.10)$$

*Proof.* Consider an identity (2.7) from theorem 3 once more, i.e.

$$-\int_0^\infty \omega^2 V_1(\omega)\cos\omega\tau d\omega - a\int_0^\infty \omega V_1(\omega)\sin\omega\tau\, d\omega + b\int_0^\infty V_1(\omega)\cos\omega\tau\, d\omega \equiv 0 \qquad (2.7)$$

and substitute $\omega = \frac{\omega_2}{-\alpha}$ ($\alpha > 0$) in the function

$$V_1(\omega) = \frac{1}{\pi}\frac{2\chi(\chi^2 + \lambda^2)}{[\chi^2 + (\lambda + \omega)^2][\chi^2 + (\lambda - \omega)^2]} = \frac{2\chi(\chi^2 + \lambda^2)}{\pi[\chi^2 + (\lambda + \frac{\omega_2}{-\alpha})^2][\chi^2 + (\lambda - \frac{\omega_2}{-\alpha})^2]} =$$

$$= \frac{2\chi(\chi^2 + \lambda^2)}{\pi[\frac{(-\alpha\chi)^2}{(-\alpha)^2} + (\frac{-\alpha\lambda}{-\alpha} + \frac{\omega_2}{-\alpha})^2][\frac{(-\alpha\chi)^2}{(-\alpha)^2} + (\frac{-\alpha\lambda}{-\alpha} - \frac{\omega_2}{-\alpha})^2]} =$$

$$= -\alpha\frac{2\chi_2(\chi_2^2 + \lambda_2^2)}{\pi[\chi_2^2 + (\lambda_2 + \omega_2)^2][\chi_2^2 + (\lambda_2 - \omega_2)^2]} = \overline{\overline{V}}_1(\omega_2),$$

where $\chi_2 = -\alpha\chi$, $\lambda_2 = -\alpha\lambda$.

After substitution in the identity (2.7) we note that $\omega_2 \in (-\infty, 0]$, since $\omega \in [0, \infty)$:

$$-\int_0^\infty \omega^2 V_1(\omega)\cos\omega\tau d\omega - a\int_0^\infty \omega V_1(\omega)\sin\omega\tau\, d\omega + b\int_0^\infty V_1(\omega)\cos\omega\tau\, d\omega \equiv$$

$$\equiv -\frac{1}{-\alpha^3}\int_0^{-\infty} \omega_2^2 \overline{\overline{V}}_1(\omega_2) \cos\frac{\omega_2}{-\alpha}\tau\, d\omega_2 - \frac{1}{\alpha^2}a\int_0^{-\infty} \omega_2 \overline{\overline{V}}_1(\omega_2)\sin\frac{\omega_2}{-\alpha}\tau\, d\omega_2 - \frac{1}{\alpha}b\int_0^{-\infty} \overline{\overline{V}}_1(\omega_2) \cos\frac{\omega_2}{-\alpha}\tau d\omega_2 \equiv$$

$$\equiv -\int_{-\infty}^0 \omega_2^2 \overline{\overline{V}}_1(\omega_2)\cos\frac{\omega_2}{-\alpha}\tau\, d\omega_2 + \alpha a\int_{-\infty}^0 \omega_2 \overline{\overline{V}}_1(\omega_2)\sin\frac{\omega_2}{-\alpha}\tau\, d\omega_2 + \alpha^2 b\int_{-\infty}^0 \overline{\overline{V}}_1(\omega_2) \cos\frac{\omega_2}{-\alpha}\tau d\omega_2 \equiv 0. \quad (2.11)$$

In the identity (2.11):

a) $\dfrac{\omega_2}{-\alpha} > 0$;

b) the sign has changed before the coefficient a is compared to (2.7).

The substitution resulted in the following changes:

– in the function $\overline{\overline{V}}_1(\omega_2)$ we have parameters

$$\chi_2 = -\alpha\chi, \qquad \lambda_2 = -\alpha\lambda;$$

– a characteristic equation corresponding to the differential equation (2.2), prior to the substitution was

$$\eta^2 + \alpha a\eta + \alpha^2 b = 0,$$

while, after the substitution it became

$$\eta^2 - \alpha a\eta + \alpha^2 b = 0.$$

Consequently, the variable substitution resulted in changed values of coefficients a and b in the equation (2.2) and they are now equal to $\alpha a$ and $\alpha^2 b$, the sign before the coefficient a has also changed. These changes are not possible since coefficients in a differential equation are specified. Consequently, after the sign changes the argument $\tau$ в $v_1(\tau)$ should also change so that in (2.2) the coefficients remain unchanged. Hence, taking into account the notation in $\overline{\overline{V}}_1(\omega_2)$ we obtain

$$v_1(\tau) = (\cos\lambda\tau + \dfrac{\chi}{\lambda}\sin\lambda\tau)\exp(-\chi\tau) = [\cos(-\alpha\lambda)(\dfrac{\tau}{-\alpha}) + \dfrac{-\alpha\chi}{-\alpha\lambda}\sin(-\alpha\lambda)(\dfrac{\tau}{-\alpha})]\exp[(\alpha\chi)(\dfrac{\tau}{-\alpha})] =$$

$$= (\cos\lambda_2\tau_2 + \dfrac{\chi_2}{\lambda_2}\sin\lambda_2\tau_2)\exp(-\chi_2\tau_2) = \overline{\overline{v}}_1(\tau_2),$$

where

$$\tau_2 = \dfrac{\tau}{-\alpha} < 0,$$

since $\tau > 0$. The product of frequency $\omega$ and argument $\tau$ does not change

$$\omega\tau = \dfrac{\omega_2}{-\alpha}(-\alpha\tau_2) = \omega_2\tau_2. \qquad (\omega_2 < 0,\ \tau_2 < 0) \qquad (2.12)$$

Taking into account the relationship (2.12) the function $\sin\omega\tau$ behaves in

$$\dfrac{dv_1(\tau)}{d\tau} = \int\limits_0^\infty \omega V_1(\omega)\sin\omega\tau\, d\omega$$

as an even function after expanding integration limits in this integral and as a result the integrand is odd. We can use that result in the following way. Create the function

$$v_m(\tau_1) = \dfrac{1}{2}\int\limits_{-\infty}^\infty [\delta_1(\Omega - \omega) + \delta_2(\Omega + \omega)]\, \overline{V}_1(\Omega)\cos\Omega\tau_1 d\Omega$$

where $\delta_1(\Omega - \omega)$ and $\delta_2(\Omega + \omega)$ are Dirac's functions and note, that the first derivative of this function equals zero due to (2.12):

$$\frac{dv_m(\tau_1)}{d\tau_1} = -\frac{1}{2}\int_{-\infty}^{\infty} [\delta_1(\Omega - \omega) + \delta_2(\Omega + \omega)]\Omega\, \overline{V}_1(\Omega) \sin\Omega\tau_1\, d\Omega =$$

$$= -\frac{1}{2}\{\omega\, \overline{V}_1(\omega) \sin\omega\tau_1 + (-\omega) \sin[(-\omega)(-\tau_1)]\, \overline{V}_1(-\omega)\} = -\frac{1}{2}\{\omega\, \overline{V}_1(\omega) \sin\omega\tau_1 - \omega\, \overline{V}_1(\omega) \sin\omega\tau_1\} \equiv 0.$$

For the functions $v_m(\tau_1)$ и $\dfrac{d^2 v_m(\tau_1)}{d\tau_1^2}$ we obtain:

$$v_m(\tau_1) = \frac{1}{2}\int_{-\infty}^{\infty} [\delta(\Omega - \omega) + \delta(\Omega + \omega)]\, \overline{V}_1(\Omega) \cos\Omega\tau_1\, d\Omega = \overline{V}_1(\omega) \cos\omega\tau_1,$$

$$\frac{d^2 v_m(\tau_1)}{d\tau_1^2} = -\frac{1}{2}\int_{-\infty}^{\infty} [\delta(\Omega - \omega) + \delta(\Omega + \omega)]\Omega^2\, \overline{V}_1(\Omega) \cos\Omega\tau_1\, d\Omega = -\omega^2\, \overline{V}_1(\omega) \cos\omega\tau_1.$$

Substituting the function $v_m(\tau_1)$ with its derivatives in the equation (2.4) we obtain

$$-\omega^2\, \overline{V}_1(\omega) \cos\omega\tau_1 + \alpha^2 b\, \overline{V}_1(\omega) \cos\omega\tau_1 = 0,$$

when

$$\omega^2 = \alpha^2 b,$$

i.e. the frequency $\omega$, the coefficient b in (2.4) and the arbitrary parameter $\alpha$ are linked via the relationship (2.10).

The theorem 4 is proved when we have i = 1 in the identity (2.6).

The prove of this theorem is similar for i = 2, 3 n the identity (2.6).

The theorem 4 is proved fully.

### 3. The Universe structure.

The Universe structure will be discussed as applied to a specific differential equation

$$[T^2 p^2 + 2\xi T p + 1]^3 y + [T_1^2 p^2 + 1]^2 \varphi(y) = 0, \quad \varphi(0) = 0, \quad (0 < \xi < 1) \qquad (3.1)$$

where $p = \dfrac{d}{dt}$ is the differentiation operator, $T > 0$ and $T_1 > 0$ are the constants ($T \neq T_1$). Assume that (3.1) is asymptotically stable for arbitrary initial conditions and arbitrary nonlinear function

$$0 \leq y\varphi(y) \leq ky^2.$$

In this statement the problem is known in the theory of motion stability as the problem of absolute stability. Since $\varphi(y)$ is arbitrary, we will treat cases when $\varphi(y) = 0$ and $\varphi(y) = ky$, which present a special interest, since correlation functions according to theorem 1 can be obtained only for a linear equation

$$[T^2 p^2 + 2\xi T p + 1]^3 y + k[T_1^2 p^2 + 1]^2 y = 0, \quad (0 < \xi < 1) \qquad (3.2)$$

in which the parameter k, as shown below, can assume values from $[0, k_{cr}]$.

Note, that theoretical results are in no way linked to this equation, however, its use facilitates considerably further explanation, as becomes obvious from what follows.

According to Newton, motion in the Universe is described by three equations of the second order, while (3.2) is an equation of the sixth order, i.e. using the roots (3.2) one can always obtain the three equations required.

Any equation which describes motion of material bodies, also in the Universe, is nonlinear, for the Universe, this equation should also be asymptotically stable which is confirmed by many years of astronomic observations. Hence, the author assumes that this motion can be described qualitatively as (3.1), and it is required to obtain the value of the maximal k, when this linear equation (3.2) is asymptotically stable, since only then correlation functions according to theorem 1 can be obtained. We will come back to the discussion of this issue in what follows, for now we note that for k = 0 we obtain an equation for the case when there are no material bodies. In this case in a three-dimensional space any of the second-order equations should be the same since not a single Cartesian coordinate has any advantages a priori. In equation (3.2) this condition is satisfied.

Consider a case when $\varphi(y) = 0$ in (3.1) or, which is the same, k = 0 in (3.2):

$$[T^2 p^2 + 2\xi T p + 1]^3 y = 0. \quad (0 < \xi < 1) \quad (3.3)$$

Roots in this equation are complex-conjugated, let us denote their parameters as $\chi \pm i\lambda$ (for convenience of reading all designations are identical to the ones used in theorem 1). Along with the solution

$$y_1(t) = (\cos\lambda t + \frac{\chi}{\lambda} \sin\lambda t)\exp(-\chi t),$$

obtained by using a pair of complex-conjugated roots of this equation and discussed for $t \geq 0$, we obtain in accordance with theorem 1 a solution as well

$$v_1(\tau) = 2\int_0^\infty V_1(\omega)\cos\omega\tau \, d\omega, \quad (3.4)$$

where $-\infty < \tau < \infty$.

Function $v_1(\tau)$ by virtue of theorem 1 is an auto-correlation function of a stationary random process $\xi(t)$. The solution (3.4) is substituted in equations (3.3) and, after differentiating under the sign of integration with respect to $\tau$ as a parameter, we obtain identity

$$-T^2\int_0^\infty \omega^2 V_1(\omega)\cos\omega\tau d\omega - 2\xi T\int_0^\infty \omega V_1(\omega)\sin\omega\tau d\omega + \int_0^\infty V_1(\omega)\cos\omega\tau d\omega \equiv 0.$$

According to theorem 3 as a result of substitution of the variable $\omega = \frac{\omega_1}{\alpha}$ ($\alpha > 0$) in this identity we can obtain a new parametric family of the equations

$$T^2 \frac{d^2v_1}{d\tau_1^2} + 2\xi\alpha T \frac{dv_1}{d\tau_1} + \alpha^2 v_1 = 0, \qquad (3.5)$$

$$\tau_1 = \frac{\tau}{\alpha},$$

and a new family of the identities

$$-T^2 \int_0^\infty \omega^2 \overline{V}_1(\omega)\cos\omega\tau_1 d\omega - 2\alpha\xi T \int_0^\infty \omega \overline{V}_1(\omega)\sin\omega\tau_1 d\omega + \alpha^2 \int_0^\infty \overline{V}_1(\omega)\cos\omega\tau_1 d\omega \equiv 0, \qquad (3.6)$$

where

$$\overline{V}_1(\omega) = \frac{1}{\pi} \int_0^\infty \overline{v}_1(\tau_1)\cos\omega\tau_1 \, d\tau_1,$$

$$v_1(\tau) = (\cos\lambda\tau + \frac{\chi}{\lambda}\sin\lambda\tau)\exp(-\chi\tau) = (\cos\alpha\lambda\frac{\tau}{\alpha} + \frac{\chi}{\lambda}\sin\alpha\lambda\frac{\tau}{\alpha})\exp(-\alpha\chi\frac{\tau}{\alpha}) =$$

$$= (\cos\alpha\lambda\tau_1 + \frac{\alpha\chi}{\alpha\lambda}\sin\alpha\lambda\tau_1)\exp(-\alpha\chi\tau_1) = \overline{v}_1(\tau_1). \quad (\tau_1 \geq 0) \qquad (3.7)$$

According to theorem 4 expanding the integration limits by $\omega$ in (3.6) we obtain a new solytion

$$v_m(\tau_1) = \frac{1}{2} \int_{-\infty}^\infty [\delta_1(\Omega - \omega) + \delta_2(\Omega + \omega)] \overline{V}_1(\Omega)\cos\Omega\tau_1 d\Omega, \qquad (3.8)$$

in the equation (3.5), if the frequency $\omega$, the arbitrary parameter $\alpha$ and the coefficient T in (3.5) are connected by the relationship

$$\frac{\omega}{\alpha} = \pm\frac{1}{T}.$$

The solution (3.8) reduces the parametric family of the equations (3.5) to the another parametric family of the equations, since for any pair frequency $\pm\omega$ we obtain

$$\frac{dv_m(\tau_1)}{d\tau_1} \equiv 0. \qquad (3.9)$$

The solution $v_m(\tau_1)$ is in fact a monochromatic oscillation $\cos\Omega\tau_1$, in which $\Omega$ is a random variable of a new type:

a) according to the probability theory $\Omega$ is a discrete random value, since possible values of $\Omega = \pm\omega$, where all frequencies in pairs from the spectrum $\overline{V}_1(\omega)$, except for $\omega = 0$;

b) possible values of $\Omega$ are all frequency values from the spectrum of the probability density $\overline{V}_1(\omega)$, i.e. $\Omega$ is a continuous random value.

The solution (3.8) represents harmonic oscillations of any frequencies which can be treated as waves, when $\Omega$ is a continuous value. However, they can be also treated as particles, when $\Omega$ is a discrete value. Note that we discussed the equation (3.2) for k = 0, i.e. equation of motion in the absence of material bodies in the Universe. Today it is known that harmonic oscillations characterized by corpuscular-wave dualism are called light. Besides, photons indeed do not have mass at rest. This leads to a statement that (3.5), (3.8) and (3.9) are the equation, the solution and the property of white light accordingly.

It has to be noted that we also have here substantial differences with what is known in physics:

a) harmonic oscillations which we identify as light are a correlation function of a stationary random process $\xi(t)$;

b) an argument in these oscillations is

$$\tau = t_2 - t_1$$

a time interval between sections $\xi(t)$ at arbitrary times $t_1$ и $t_2$. Since light and electromagnemtic wave are of the same nature, we can state almost surely that an argument in Maxwell's equations should also be $\tau$, i.e. electromagnetic wave equations and light equations are equations of the electromagnetic field with the same argument $\tau$. In its turn electromagnetic field is a feature of stationary and Euclidean Universe.

To obtain correlation functions using roots of equation (3.2) for any $k \neq 0$, we must obtain the maximal value of k, when the real-valued part of the roots is negative which is necessary for obtaining this function according to theorem 1. This problem can be solved using the Nyquist criterion, and to do that (3.2) is to be treated as a system with input ky and output –y, i.e.

$$[T^2 p^2 + 2\xi T p + 1]^3 y_{out} = [T_1^2 p^2 + 1]^2 y_{in}$$

and study its frequency characteristic from $y_{in}$ to $y_{out}$:

$$W(i\omega) = \frac{[T_1^2 (i\omega)^2 + 1]^2}{[T^2 (i\omega)^3 + 2\xi T(i\omega)^2 + 1]^3} . \qquad (0 < \xi < 1)$$

Figure 1 shows the frequency characteristic of an «open» system $W^*(i\omega)$, which corresponds to the case k = 0:

$$W^*(i\omega) = \frac{1}{[T^2 (i\omega)^3 + 2\xi T(i\omega)^2 + 1]^3} .$$

Figure 2 shows the frequency characteristic of the closed system. The constant $T_1$ in $W(i\omega)$ is chosen to meet the condition $T_1 \omega_0 = 1$, where $\omega_0$ is a frequency, at which $\text{Im} W^*(i\omega_0) = 0$, $\text{Re} W^*(i\omega_0) < 0$, and the constant T is chosen to meet the condition

$$3 \arg (1 - T^2 \omega_0^2 + 2\xi T i \omega_0) = \pi.$$

The characteristic $W(i\omega)$ starts on a positive real-valued semi-axis of the complex plane W and runs six quadrants sequentially. According to Nyquist the roots in (3.2) have negative real-valued parts for $k \in [0, \infty)$, since the frequency characteristic does not cross the negative real-valued axis.

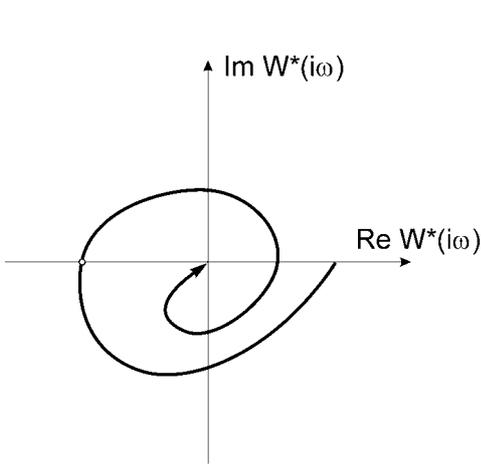 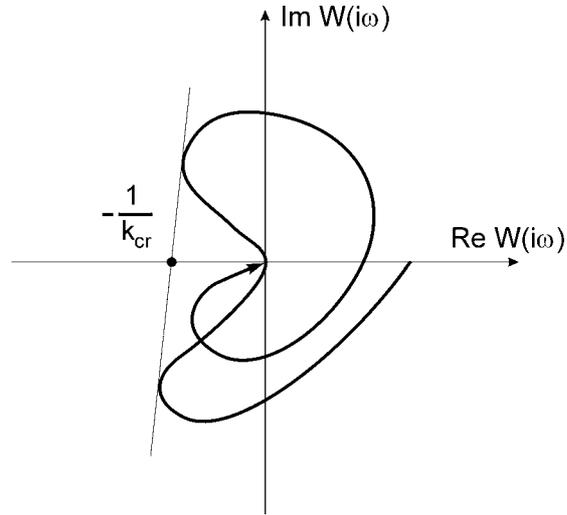

Fig. 1.  Fig. 2.

Since we discuss equation (3.2) in the parameter space, then k is treated as «distance» of the Euclidean space.

There is no analytical method to obtain the roots of equation (3.2) for an arbitrary $k \neq 0$. Hence they were obtained by modeling. For an arbitrary k roots in (3.2) are complex-conjugated. When $k \to \infty$ we can write (3.2) as

$$\lim_{k \to \infty} \frac{1}{k}[T^2 p^2 + 2\xi Tp + 1]^3 y + [T_1^2 p^2 + 1]^2 y = 0,$$

whence it follows that we have an equation

$$[T_1^2 p^2 + 1]^2 y = 0.$$

Consequently, out of two pairs of roots the real-valued part tends to zero, and the imaginary part tends to a concrete value

$$\omega = \pm \frac{1}{T_1},$$

which can be obtained using the frequency characteristic (it is a frequency at which $W(i\omega)$ crosses the origin of coordinates). The modeling results facilitate a lot the exercise since correlation functions are to be obtained for only one type of roots in accordance with the theorem 1

Consider equation (3.2) and find out up to what value of k correlation functions can be obtained according to theorem 1. The correlation function is a positive definited function [5]. Consequently, the trivial solution of the equation (3.2) must be the positive definited function too. It imposes for the roots of the equation (3.2) more stringent conditions than just negative real part. Without going into the details we only note that the correlation function can be obtained for $k \in [0, k_{cr}]$, when the complex plane W can be divided by a straight line into two parts so that the whole of $W(i\omega)$ lies strictly to the right of this line, and the abscissa $-\frac{1}{k_{cr}}$ of the limit line is minimally separated from the origin of the coordinates [3]. This straight line and the value of $k_{cr}$ are shown in Fig. 2. This implies that for an arbitrary k we have a light equation up $k = k_{cr}$.

Boundary of asymptotic motion stability, according to the nonlinear equation obviously cannot be more than $k_{cr}$. It can be proved that within a galaxy the stability boundary of the nonlinear equation (3.1) coincides with the light equation boundary in the linear equation.

Parameters of complex-conjugated roots of equation (3.2) when k changes to $k = k_{cr}$ are not known to us analytically, however, white light equations with their solution and the features of that solution are going to be similar to that of (3.5), (3.8) and (3.9), obtained for k = 0. Moreover, the light equation we will derive from the equations describing the motion of the Solar system, and the Earth is part of the System.

*Experiments by Mickelson – Morley in 1887 can be treated as the experimental confirmation of the fact that light is ether which fills the space, hence it is not possible to detect the motion of the Earth through ether with the help of light.*

Correlation functions obtained for various values of k in (3.2) will behave similar to the function $v_1(\tau)$ in (3.7). This implies that the argument $\tau$ will "narrow down" or "expand" with the help of an arbitrary $\alpha$. Variation of $\tau$ in the function $v_1(\tau)$ results in a similar change of the argument t in the function $y_1(t)$. In an ordinary differential equation (3.3) no transformation of (3.7) is done, since, first, this leads to a change of an independent argument t: if t does not depend on us, then we should not try change it. Second, and this is not acceptable either, (3.7) changes coefficients of this equation, determined by the Newton's laws. However, it is one thing when (3.7) is done artificially, but it is quite different when such transformation is a result of the integration variable substitution. Let us consider this problem when the complex-conjugated root in the equation loses its real-valued part ($\chi = 0$). This situation can be obtained from theorem 3, when in the function

$$V_1^*(\omega) = \alpha \frac{2\chi_1(\chi_1^2 + \lambda_1^2)}{\pi[\chi_1^2 + (\lambda_1 + \omega_1)^2][\chi_1^2 + (\lambda_1 - \omega_1)^2]}, \quad \chi_1 = \alpha\chi, \lambda_1 = \alpha\lambda,$$

$\chi_1 \to 0$. One can easily see that the numerator will be zero. The denominator, however, will also be zero since for an arbitrary $\alpha$ there is always a frequency when $\lambda_1 = \alpha\lambda = \pm\omega_1$. Consequently, finding the value of an indeterminate of the type $\frac{0}{0}$, we obtain $\delta$–functions which are easy to integrate. As a result, we obtain an above family of equations (3.5) with the only difference being that for $\chi_1 \neq 0$ the first derivative coefficient was by-passed due to (3.9), and now the coefficient will just be equal to zero. Let us write down this family of equations in a general form (not following the example of (3.5), since the case when $\chi_1 = 0$ is not possible), i.e.

$$\frac{d^2v_1}{d\tau_1^2} + \alpha^2 bv_1 = 0 . \qquad (3.10)$$

Compare (3.10) witm the well-known mathematical problem of celestial mechanics of movement of two material bodies with masses M and m in the Euclidean space:

$$\frac{d^2x}{dt^2} = -\frac{\mu x}{r^3}, \qquad \frac{d^2y}{dt^2} = -\frac{\mu y}{r^3}, \qquad \frac{d^2z}{dt^2} = -\frac{\mu z}{r^3},$$

where $\mu$ is a constant value which is a function of masses M and m, while

$$r = \sqrt{x^2 + y^2 + z^2}$$

is a distance between the moving object and the origin of coordinates. These equations are particular cases of the equation (3.10), when

$$\alpha^2 b = \frac{\mu}{r^3}, \qquad t = \tau$$

and there are no masses of any moving bodies. Consequently, I. Newton, having developed the differential calculus ascribed to an arbitrary coefficient $\alpha^2 b$ a specific value, based on the observations by Kepler. This implies that the Newton's law of universal gravitation is only a model describing the motion of material bodies in a stationary environment. Note that a priority frame of reference for Newton's equations in a stationary environment can be any system of coordinates established anywhere. Here is what an eminent physicist C.Lanczos [6] wrote about it: «What is after all, a "frame of reference" ? Coould it be more than a man-made artifice, an axilliary tool we have manufactured in orger to describe the physical universe? Why should we not have the right to relate our measurements to any arbitrary frame of axes we may choose? These axes do not belong to the objective world order. We set up our axes as a kind of scaffold in order to walk on it and be able to reach any part of the building. Yet we must be aware of the fact that this scaffold is not part of building. It can be put up and pulled down without altering the building in the least. But in Newton's physics the pecular thing happened that an absolute frame was designated (not entirely, since a stricly uniformmmmm moution relative to it was tolerand) as the only correct frame to which our measurements should be related». Stationary environment can be compared to an ocean when it is completely calm. This does not mean, however, that there is no motion in a ocean. On the contrary, an ocean at any moment in time will always have waves of all frequencies. In the ocean we can establish an arbitrary system of coordinates described by Newton when introducing the notion of absolute space. This system of coordinates does not have anything to do with the stationary environment. It is just a convenient frame of reference for our calculations. We can easily dispose of them since they are not part of our ocean. This has become possible due to discovery of a new and important physical property of the solution of Newton's equations, it offers an absolutely novel angle from which to view the processes taking place in the Universe.

A stationary random process $\xi(t)$ in our case can be presented as harmonic oscillations

$$\sigma(t) = \overline{\alpha} \cos\omega t + \overline{\beta} \sin\omega t,$$

with $\omega \in (0, \infty)$, and random values $\overline{\alpha}$ and $\overline{\beta}$ described as follows:

$$E\overline{\alpha} = E\overline{\beta} = 0, \ E\overline{\alpha}\,\overline{\beta} = 0 \text{ и } E\overline{\alpha}^2 = E\overline{\beta}^2 = \overline{b}.$$

Function $\sigma(t)$, similar to its correlation function $K(\tau) = \overline{b}\cos\omega\tau$, is a solution of equation (3.5), since the change of the sign of the frequency $\omega$ also changes the sign of time t, hence, similar to (3.9),

$$\frac{d\sigma(t)}{dt} \equiv 0.$$

Consider now the notion of absolute time introduced by Newton. This is what he wrote about it: «Absolute, True and Mathematical Time, of itself and from its nature flows equably without regard to any thing external, and by another name is called Duration...». Let us prove, that t does not have this property and to do so we will write equality (2.5) from theorem 3 as follows

$$\omega\tau = \frac{\omega_1}{\alpha}\alpha\tau_1 = \omega_1\tau_1 = C_1, \qquad (3.11)$$

where $C_1$ depends on numeric values of $\omega$ and $\tau$, and this $C_1$ can be specified in advance since $\omega$ and $\tau$ are arbitrary. Since $t_1$ and $t_2$ are arbitrary in

$$\tau = t_2 - t_1$$

changes in $\tau$ in (3.11) with the use of $\alpha$ lead to similar changes of $t_1$ and $t_2$, i.e.

$$\alpha\tau = \alpha(t_1 - t_2) = \alpha t_1 - \alpha t_2,$$

and, consequently, to «expansion» or «narrowing down» of time t. Hence, it immediately follows that t does not have the property of being independent which Newton ascribed to it.

The Universe, however, has not only our Galaxy but other galaxies too, and all of them co-exist in the same space. Consider a case when the light from a galaxy arrives at our Galaxy, and, consequently, to the Solar system. It is assumed that motion of any galaxy is described by an equation of type (3.2), however, with different constant values $T_1$ and $T$, hence, different galaxies have different values of $k_{cr}$. Let us denote: for our Galaxy $k_{cr} = k_G$, for any other galaxy $k_{cr} = k_g$ and consider the equation of motion of our Galaxy (3.2) for $k = k_{cr} = k_G$. Let us obtain a second-order equation

$$\frac{d^2 Z_1}{d\tau^2} + a\frac{dZ_1}{d\tau} + bZ_1 = 0$$

with respect to a complex function

$$Z_1(\tau) = [T_1^2 p^2 + 1]^2 v_1(\tau) \qquad (3.12)$$

Note, that in accordance with theorem 3, the coefficient a for any pair of the frequence $\pm\omega$ can be "by-passed", and the coefficient b must be multiple to $\alpha^2$, i.e.. we can consider a second-order equation

$$\frac{d^2 Z_1}{d\tau^2} + \alpha^2 b Z_1 = 0. \qquad (3.13)$$

Assume that we have an equality

$$\alpha^2 b = k_G \qquad (3.14)$$

and note that this condition can always be fulfilled by the respective choice of the scale for the argument τ in accordance with theorem 3. The function (3.12) is substituted in (3.13), and we obtain the identity

$$\frac{d^2 Z_1(\tau)}{d\tau^2} + k_G Z_1(\tau) \equiv 0, \qquad (3.15)$$

which takes into account (3.14). The motion for the Galaxy (3.2) we write as

$$[T^2 p^2 + 2\xi T p + 1]^3 v_1(\tau) + k_G Z_1(\tau) \equiv 0, \qquad (3.16)$$

hence, it follows that (3.16) can be treated as a second-order equation (3.13), since

$$[T^2 p^2 + 2\xi T p + 1]^3 v_1(\tau) \equiv \frac{d^2 Z_1(\tau)}{d\tau^2}.$$

For that other galaxy we have an identity as (3.16), i.e

$$[T^{*2} p^2 + 2\xi T^* p + 1]^3 v^*_1(\tau) + k_g Z^*_1(\tau) \equiv 0,$$

in which parameters of the differential equation (3.2) differ from parameters of the Galaxy's equation, this is why they are marked with asterisk. For the same reason the corresponding solutions are also marked with asterisk. For this identity we can write a second-order identity, however, with an arbitrary α according to theorem 3, i.e.

$$\frac{d^2 Z^*_1(\tau)}{d\tau^2} + \alpha^2 b^* Z^*_1(\tau) \equiv 0.$$

Since $k_g$ can take any value let us assume that $k_g < k_G$. This value of $k_g$ can be "reduced" to $k_G$ using the parameter α:

$$\frac{d^2 Z^*_1(\tau)}{d\tau^2} + k_G Z^*_1(\tau) \equiv 0, \qquad \alpha^2 b^* = k_G.$$

As already mentioned above with respect to roots of equation (3.2), the rise of k results in the lowering of the parameter χ, which in the limit tends to zero with the parameter k tending to infinity. Hence, to "reduce" $k_g$ to $k_G$ (for learning we must compare) we need in accordance with theorem 3 take α < 1 for $k_g < k_G$. After the "reduction" the solutions of the original and the "reduced" equations in a general case obviously differ, due to scale changes that roots and arguments have to undergo. However, since both equations are equations of light and the systems of rotation described by these equations are in the same space, then light from a galaxy gets into the Galaxy, and hence, to the Solar system too. If α < 1, then the process of "reduction" leads to lower frequency and color shift towards the red in the Doppler spectrum. This may lead to an erroneous conclusion that Universe accelerates, and the lower $k_g$, the faster, since in that case there is more reddening. In a case when $k_g > k_G$, to the contrary, the process of "reduction" leads to a greater frequency and the Doppler shift towards the blue. This can be erroneously interpreted as a galaxy rushing towards us, and the larger $k_g$ the faster.

Summary

The above results show that the language of the Universe physics differs from that of the physics of the Earth. The latter is based on non stationary processes, while the physics of the Universe relies on stationary processes. Hence, the study of the Universe has to be approached differently from the way it is done.

Annex

Auxiliary propositions

*Statement* 1. If α > 0, then

$$\int_0^\infty \exp(-\alpha t)\cos\omega t\, dt = \frac{\alpha}{\alpha^2 + \omega^2}, \quad (A.1)$$

$$\int_0^\infty \exp(-\alpha t)\sin\omega t\, dt = \frac{\omega}{\alpha^2 + \omega^2}. \quad (A.2)$$

*Proof* of statement 1. Calculate the integral

$$\int_0^\infty \exp(-\alpha t + i\omega t)\, dt = \frac{1}{-\alpha + i\omega}\int_0^\infty d[\exp(-\alpha t + i\omega t)] =$$

$$= -\frac{1}{-\alpha + i\omega} = \frac{\alpha}{\alpha^2 + \omega^2} + i\frac{\omega}{\alpha^2 + \omega^2}. \quad (A.3)$$

Now note that the first integral in statement 1 is a real part (A.3), i.e.

$$\int_0^\infty \exp(-\alpha t)\cos\omega t\, dt = \text{Re}\{\int_0^\infty \exp(-\alpha t + i\omega t)dt\} = \frac{\alpha}{\alpha^2 + \omega^2},$$

While the second integral in statement 1 is an imaginary part (A.3), i.e.

$$\int_0^\infty \exp(-\alpha t)\sin\omega t\, dt = \text{Im}\{\int_0^\infty \exp(-\alpha t + i\omega t)dt\} = \frac{\omega}{\alpha^2 + \omega^2}.$$

Statement 1 is proved.

*Statement 2.* If $\alpha > 0$, then

$$\int_0^\infty \exp(-\alpha t)\cos\beta t\cos\omega t\, dt = \frac{\alpha[\alpha^2 + \beta^2 + \omega^2]}{[\alpha^2 + (\beta+\omega)^2][\alpha^2 + (\beta-\omega)^2]}, \quad (A.4)$$

$$\int_0^\infty \exp(-\alpha t)\sin\beta t\cos\omega t\, dt = \frac{\beta(\alpha^2 + \beta^2) - \beta\omega^2}{[\alpha^2 + (\beta+\omega)^2][\alpha^2 + (\beta-\omega)^2]}. \quad (A.5)$$

*Proof* of statement 2. Calculate the integral

$$\int_0^\infty \exp[-\alpha t + i(\beta \pm \omega)t]\, dt = \frac{1}{-\alpha + i(\beta \pm \omega)} \int_0^\infty d\{\exp[-\alpha t + i(\beta \pm \omega)t]\} =$$

$$= -\frac{1}{-\alpha + i(\beta \pm \omega)} = \frac{\alpha + i(\beta \pm \omega)}{\alpha^2 + (\beta \pm \omega)^2}. \quad (A.6)$$

Now note that integral in statement 2 can be easily calculated using (A.6).
First integral (A.4):

$$\int_0^\infty \exp(-\alpha t)\cos\beta t\cos\omega t\, dt = \frac{1}{2}\text{Re}\{\int_0^\infty [\exp(-\alpha t + i\beta t)][\exp(i\omega t) + \exp(-i\omega t)]dt\} =$$

$$= \frac{1}{2}\{\frac{\alpha}{\alpha^2 + (\beta+\omega)^2} + \frac{\alpha}{\alpha^2 + (\beta-\omega)^2}\} = \frac{\alpha[\alpha^2 + \beta^2 + \omega^2]}{[\alpha^2 + (\beta+\omega)^2][\alpha^2 + (\beta-\omega)^2]}.$$

Second integral (A.5):

$$\int_0^\infty \exp(-\alpha t)\sin\beta t\cos\omega t\, dt = \frac{1}{2}\text{Im}\{\int_0^\infty [\exp(-\alpha t + i\beta t)][\exp(i\omega t) + \exp(-i\omega t)]dt\} =$$

$$= \frac{1}{2}\{\frac{\beta+\omega}{\alpha^2 + (\beta+\omega)^2} + \frac{\beta-\omega}{\alpha^2 + (\beta-\omega)^2}\} =$$

$$= \frac{1}{2}\{\frac{(\beta+\omega)[\alpha^2 + (\beta-\omega)^2]}{[\alpha^2 + (\beta+\omega)^2][\alpha^2 + (\beta-\omega)^2]} + \frac{(\beta-\omega)[\alpha^2 + (\beta+\omega)^2]}{[\alpha^2 + (\beta+\omega)^2][\alpha^2 + (\beta-\omega)^2]}\} =$$

$$= \frac{(\beta+\omega)(\alpha^2 + \beta^2 - 2\beta\omega + \omega^2) + (\beta-\omega)(\alpha^2 + \beta^2 + 2\beta\omega + \omega^2)}{[\alpha^2 + (\beta+\omega)^2][\alpha_2 + (\beta-\omega)^2]} =$$

$$= \frac{\beta(\alpha^2 + \beta^2) - \beta\omega^2}{[\alpha^2 + (\beta+\omega)^2][\alpha^2 + (\beta-\omega)^2]}.$$

Statement 2 is proved.

*Statement* 3. If $\alpha > 0$, then

$$\int_0^\infty t\exp(-\alpha t)\cos\omega t\, dt = \frac{\alpha^2 - \omega^2}{(\alpha^2 + \omega^2)^2}. \tag{A.7}$$

*Proof.* This integral can be easily as

$$\int_0^\infty t\exp(-\alpha t)\cos\omega t\, dt = \int_0^\infty \frac{\partial}{\partial \omega}[\exp(-\alpha t)\sin\omega t]\, dt.$$

Now, using the value of the integral (A.2) from statement 1, one can use the theorem of differentiation with respect to the parameter:

$$\int_0^\infty \frac{\partial}{\partial \omega}[\exp(-\alpha t)\sin\omega t]\, dt = \frac{d}{d\omega}\{\int_0^\infty \exp(-\alpha t)\sin\omega t\, dt\} = \frac{d}{d\omega}\{\frac{\omega}{\alpha^2 + \omega^2}\} =$$

$$= \frac{\alpha^2 + \omega^2 - 2\omega^2}{(\alpha^2 + \omega^2)^2} = \frac{\alpha^2 - \omega^2}{(\alpha^2 + \omega^2)^2}.$$

Statement 3 is proved.